\documentclass[preprint,prd,tightenlines,nofootinbib]{revtex4}
\newcommand{\gsim}{~{}_{\textstyle\sim}^{\textstyle >}~}
\newcommand{\lsim}{~{}_{\textstyle\sim}^{\textstyle <}~}

\newcommand{\be}{\begin{equation}}
\newcommand{\ee}{\end{equation}}

\newcommand{\bea}{\begin{eqnarray}}
\newcommand{\eea}{\end{eqnarray}}

\begin{document}

\title{Nuclei as Laboratories: Nuclear Tests of Fundamental Symmetries}


\author{M. J. Ramsey-Musolf}
\address{California Institute of Technology, Pasadena, CA 91125, USA}

\begin{abstract}
The prospect of a rare isosotope accelerator facility opens up possibilities for a new generation of nuclear tests of fundamental symmetries. In this talk, I survey the current landscape of such tests and discuss future opportunities that a new facility might present.

\end{abstract}

\maketitle



\section{Introduction}

The mission of nuclear physics in the coming decade is to explain the origin, evolution, and structure of the baryonic matter in the Universe\cite{NSAC05}. Historically, nuclear physics has played a key role in developing our Standard Model (SM) of particle physics through exquisite tests of fundamental symmetries such as parity (P) and time-reversal (T) invariance. Today, we are on the cusp of a new era in particle physics, with the imminent operation of the Large Hadron Collider (LHC). One hopes that the LHC will both uncover the mechanism of electroweak symmetry-breaking associated with the yet unseen Higgs boson and discover evidence for physics beyond the SM. At the same time, cosmology points to the need for new particle physics, as the SM fails to account for the abundance of visible, baryonic matter in the Universe, the existence of cold dark matter, and the mysterious dark energy responsible for cosmic acceleration. The corresponding challenge for particle and nuclear physics is to determine what physics beyond the SM can explain these cosmic building blocks and their relative contributions to the energy density of the Universe.

Within this context, the focus for nuclear physics falls squarely on the baryonic component. It is up to nuclear physics to explain how baryonic matter  came to be in the first place; how it evolved as the Universe cooled;  how it coalesced into the protons and neutrons and other hadrons; how the fundamental quarks and gluons are put together inside baryons; how their interactions give rise to atomic nuclei in all their rich complexity; and how these nuclei and their interactions drive the formation, structure, and life cycle of stars. In pursuing this mission, it is essential to understand how the basic forces of nature have shaped this story of the baryons. Despite the tremendous successes of nuclear physics over the decades, our knowledge of this story remains quite limited. We are able to account for many -- but not all --  properties of baryonic matter only after quarks and gluons coalesced into protons and neutrons. We don't know, however, whether the strong, electroweak, and gravitational forces that dominate at late times were the only or even the most important ones at times before the formation of QCD bound states. We don't know whether the interactions of quarks with leptons looked the same at earlier times as they appear today. And we certainly have no definitive explanation of how the tiny -- but anthropically relevant -- excess of baryonic over anti-baryonic matter arose after the initial blast of the Big Bang. 

In this talk, then, I want to discuss how we can exploit atomic nuclei as \lq\lq laboratories" to look back in time to try and arrive at insights into these questions. In doing so, we have to exploit what we know about the symmetries of SM interactions and  -- by performing precise tests of these symmetries -- piece together clues about the interactions of baryonic matter at early cosmic times.
At the same time, there remain a host of poorly understood features of baryonic matter in the present era, and among the most enigmatic, is the weak interaction between quarks. So I will divide this talk into two broad questions: 
\begin{itemize}

\item[i)] {\em What were the fundamental forces that governed the interactions of quarks in the early Universe? } In particular, we would like to determine what forces were responsible for the generation of the excess of baryonic matter as well as what other forces shaped the dynamics of quarks once they were created. In addressing this question, I will discuss how searches for the permanent electric dipole moments (EDMs) of the neutron and atoms and precision studies of weak nuclear decays can provide important insights.  In each case, the violation of fundamental symmetries -- including P, T, and C -- provide essential handles in probing the forces of the early Universe.

\item[ii)] {\em How do the electroweak and strong interactions of the SM shape the weak interactions of baryons in the present Universe?} Because the strong and electromagnetic interactions are so much more powerful than the weak quark-quark interaction at the low-energies relevant to nuclear physics, we must exploit the P-violating (PV) character of the weak interaction to observe it experimentally. I will discuss how a new generation of hadronic PV experiments are poised to provide a new window on the hadronic weak interaction (HWI) and how -- ultimately -- these studies may help is better understand the nuclei as laboratories as per point (i) above.

\end{itemize}

Before proceeding, I wish to acknowledge that I will be forced to omit some important topics in nuclear physics tests of fundamental symmetries and will undoubtedly and unintentionally miss some important references to recent work.  I hope that these omissions will be remedied by other recent reviews that I have given\cite{Ramsey-Musolf:2006ur,Erler:2004cx} , and I refer the interested reader to those discussions.

\section{Electric Dipole Moments and the Origin of Baryonic Matter}

It is widely believed that the initial, post-inflationary conditions were matter-antimatter symmetric. If so, then  the particle physics of the post-inflationary era would have to be responsible for generating a nonvanishing baryon number density, $n_B$. Expressed as a ratio to the photon entropy density at freeze out, $s_\gamma$, the baryon asymmetry is
\be
Y_B\equiv \frac{n_B}{s} = 
\biggl\{
\begin{array}{cc}
(7.3\pm 2.5)\times 10^{-11}, & \text{BBN \cite{Eidelman:2004wy}}\\
(9.2\pm 1.1)\times 10^{-11}, & \text{WMAP \cite{Spergel:2003cb}}
\end{array}
\ee
where the first value (BBN) is obtained from observed light element abundances and standard Big Bang Nucleosynthesis and the second value is obtained from the cosmic microwave background as probed by the WMAP collaboration. Forty years ago, Sakharov identified the three key ingredients
for any successful accounting for this number\cite{Sakharov:1967dj}: (1) a violation of baryon number ($B$) conservation; (2) a violation of both $C$ and $CP$ symmetries; and (3) a departure from thermal equilibrium at some point during cosmic evolution\footnote{The last ingredient can be replaced by $CPT$ violation, and some baryogenesis scenarios exploit this fact.}. 

In principle, these ingredients could have generated $Y_B\not= 0$ at any moment in the post-inflationary epoch up to the era of electroweak symmetry breaking. At one extreme, baryogenesis might have occurred at very early times, associated with particle physics at scales much greater than the electroweak scale, $M_{\rm wk}$. At the other end is the possibility of electroweak baryogenesis (EWB). And, $Y_B$ may have been generated at some point between these two cosmic \lq\lq bookends".  During the coming decade, experiments that probe new weak scale physics, including EDM searches and the LHC studies, will test EWB with revolutionary power. In the most optimistic scenario, these experiments will uncover the building blocks of EWB and point to the new physics scenario that consistently incorporates them. Even null results, however, would be interesting, as they would imply that EWB is highly unlikely and point to higher scale scenarios -- such as GUT baryogenesis or leptogenesis -- that are more difficult to test experimentally.  In the case of leptogenesis one can at least look for some of the elements that have to exist to make this scenario plausible: CP-violation in the lepton sector, an appropriate scale for $m_\nu$; and lepton number violation. Neutrino oscillation studies -- together with ordinary $\beta$-decay and neutrinoless double $\beta$-decay ($0\nu\beta\beta$) -- provide our means for doing so\footnote{It should be emphasized, however, that the absence of experimental evidence for lepton C- and CP-violation does not preclude leptogenesis.}. Below, I will comment on $0\nu\beta\beta$ in more detail. 

In the case of EWB, it is natural to ask why the SM is insufficient since the SM 
contains all of Sakharov's criteria: 

\begin{itemize}

\item[(1)]{\em  Baryon number violation}.  In the SM takes place in the through anomalous processes called sphaleron transitions. The difference of baryon lepton number currents, $j_\mu^B-j_\mu^L$, is anomaly-free and conserved in the SM, but their sum is not. There exist an infinite number of vacua differing in total $B+L$ by integer units, and the probability for tunneling between these $T=0$ vacua is highly suppressed. At  temperatures of order the electroweak scale ($T_{\rm wk}$), however, the thermal excitations of gauge configurations with energy of order $T_{\rm wk}$ can occur with finite probability. These configurations can decay to a different vacuum than the one out or which they were created, thereby changing $B+L$. 

\item[(2)] {\em Departure from thermal equilibrium}. In principle, the generation of mass in the SM at the electroweak scale could have provided the necessary departure from thermal equilibrium through a transition between the phases of unbroken and broken electroweak symmetry. In order to ensure that any non-zero $n_B$ created by sphaleron processes is frozen into the broken phase (where we live), this phase transition needs to be strongly first order. The parameters of the Higgs potential are critical in determining whether or not such a first order phase transition can occur. Given the present lower bounds on the Higgs boson mass obtained from LEP II ($m_H > 114.4$ GeV) the parameters of the Higgs potential that depend on this mass and on the weak scale cannot lead to a strong first order phase transition. New physics that couples to the Higgs sector is needed to bring about such a phase transition. 

\item[(3)] {\em CP-violation}. The presence of CP-violation is needed to generate a net asymmetry between production of left- and right-handed particle densities, and it is this imbalance that feeds the $B+L$ violating sphaleron processes\footnote{This chiral charge production can be indirect, taking place first via generation of a net Higgs number density and then transferring to the fermion sector through Yukawa-like interactions.}. The electroweak sector of the SM contains CP-violation via the phase in the CKM matrix that affects interactions between the spacetime varying Higgs vev and quarks at the phase transition boundary. Unfortunately for the SM, the net effect of this CP violation is highly suppressed by the Jarlskog invariant, $J$, and multiple powers of quark Yukawa couplings\cite{Shaposhnikov:1987tw}\footnote{Farrar and Shaposhnikov subsequently argued that the SM baryon asymmetry is suppressed only by $J$ and $m_s-m_d$\cite{Farrar:1993sp}.}. Consequently, viable EWB requires the presence of new CP-violating interactions beyond those of the SM that do not suffer from this suppression. In the absence of such Jarlskog suppression, one would also expect the effects of these CP-violating interactions to enter more strongly in EDMs than does SM CP-violation. 

\end{itemize}

As reviewed in Ref.~\cite{Riotto:1999yt}, there exist a number for  models for new physics at the weak scale that can remedy these SM shortcomings. A strong, first order phase transition naturally arises in supersymmetric models, for example, as they contain new scalar degrees of freedom -- such as the scalar superpartners of fermions or singlet Higgs fields -- that couple appropriately to the SM-like Higgs and correspondingly modify the potential. These supersymmetric interactions can also include new CP-violating effects that are not suppressed as in the SM, thereby allowing for the requisite creation of chiral charge. Whether or not such models can lead to successful EWB depends on the characteristics of the Higgs potential, the details of quantum transport at the phase boundary, and on the values of the model parameters.

In addressing these issues, improvements in both theory and experiment are important. Theoretically, refined treatments of quantum transport, the details of the Higgs potential, and the dynamics of the expanding regions of broken electroweak symmetry (\lq\lq bubbles") are being pursued by our group and several others\cite{riotto}. Experimentally, searches for new physics at the LHC, precision electroweak tests, and both EDM and dark matter searches are poised to provide critical new information about the shape of any new physics at the electroweak scale. From the standpoint of EWB, the  EDM experiments will be essential in nailing down the parameters associated with new CP-violation. Moreover, experiments carried out with different systems -- such as the electron, muon, neutron, neutral atoms, and even the deuteron -- can provide complementary information. If, for example, a non-zero EDM is observed in one such system, there will exist a variety of possible models that can explain it. The results of experiments in other systems will be needed to sort out among the competing explanations. 

I would like to emphasize that the possibilities for new, neutral atom EDM searches using rare isotopes are quite compelling in this respect, and they could remain powerful probes of new CP-violation relevant to baryogenesis well into the LHC era. To illustrate, consider EWB in the Minimal Supersymmetric Standard Model (MSSM). The CP-violating phases relevant to MSSM EWB are already highly constrained by present limits obtained from EDM searches if the masses of all the supersymmetric particles participating in one-loop contributions to the EDMs are below about one TeV. It is possible to relax the EDM constraints if the masses of the sleptons and squarks (the scalar, supersymmetric partners of the leptons and quarks) are allowed to become heavy -- on the order of 3-10 TeV (see, {\em e.g.}, Ref.~\cite{Cirigliano:2006dg} and references therein). In this case, the EDMs of the electron and neutron are dominated by two-loop graphs that involve virtual, SM particles and the superpartners of the electroweak gauge and Higgs bosons -- the gauginos and Higgsinos, respectively. The situation with diamagnetic neutral atoms is somewhat different, as the neutral atom EDMs are dominated by long range, CP-violating forces the nucleus that arise from the so-called "chromoelectric" dipole moment (CDM)\cite{Pospelov:2005pr}. In the limit of heavy sfermions, the one-loop CDMs are suppressed just like the one-loop EDMs. However, there are no two-loop CDM contributions associated with virtual gauginos and Higgsinos unlike the EDM case. Thus, if nature had selected this variant of SUSY, one could expect to see non-vanishing electron and neutron EDMs that are consistent with the CP-violation needed for EWB but vanishing atomic EDMs in systems such as Xe or Ra.

\section{Weak Decays}

The study of weak decays of hadrons has an illustrious history in nuclear physics and has provided key input for the development of the SM. The classic experiment on $^{60}$Co by Wu {\em et al.} \cite{Wu:1957my}  that provided evidence for parity violation in the weak interaction -- followed by the analogous demonstration in polarized $\mu^+$ decay\cite{Garwin:1957hc} -- pointed the way to our understanding of the $(V-A)\otimes(V-A)$ structure of the SM CC interaction at low energies. Similarly, the comparison of Fermi constants extracted from muon and $\beta$-decay lead to the development of Cabibbo mixing between $\Delta S=1$ and $\Delta S=0$ charged currents and to 
notation of that weak and strong interaction eigenstates of quarks are not identical -- precursors of the full CKM model for quark mixing. Today, superallowed nuclear $\beta$-decays provide the most precise determination of  any element of the CKM matrix, namely, $V_{ud}$ -- a triumph of precision nuclear physics that has been recognized by last year's award of the Bonner Prize to Hardy and Towner. 

That history of accomplishment notwithstanding, one might ask what relevance nuclear $\beta$-decay studies will have in the coming LHC era. From my perspective, the answer is all about precision. To the extent that one can push the experimental sensitivity of various $\beta$-decay studies significantly beyond levels, this use of nuclei as laboratories will provide powerful probes of new forces at the weak scale that can complement what we may learn from the LHC or even a subsequent $e^+e^-$ collider. The way $\beta$-decay studies do so is to search for tiny deviations from SM predictions for various quantities, such as half lives or decay correlation parameters. The pattern of such deviations -- or of their absence -- can provide new clues about the character of new physics at the electroweak scale if it exists and in some cases test elements of new physics models that are more difficult to access with colliders. The key, however, is to carry out these experiments with greater degrees of precision than before -- a prospect that a new rare isotope facility, together with the new fundamental neutron physics beam line at the Spallation Neutron Source -- make possible. Let me illustrate this idea with two kinds of $\beta$-decay tests.

\vskip 0.1in

\noindent {\em $\beta$-decay and CKM unitarity.} The CKM matrix in the SM is unitarity, so any apparent departure from this unitarity property can point to new physics. The most precise such test involves elements of the first row, including $V_{ud}$ obtained from superallowed nuclear $\beta$-decays; $V_{us}$ that is obtained most precisely from kaon leptonic ($K_{e3}$) decays; and $V_{ub}$ determined from $B$-meson decays. The value of $V_{ub}$ is too small to be relevant to such tests, given the level of precision with which the other two first row elements are known, so I will focus on them. As is well-known to people at this meeting, there had been a $\gsim 2\sigma$ deviation from first row unitarity for many years, and it was long thought by many outside our field that the culprit lay in some poorly understood feature of nuclear structure. I have always found that objection hard to swallow in light of the impressive agreement between corrected $ft$ values among the various decays. If the CVC property of the SM is valid, then the corrected $ft$ values for all superallowed decays should be identical. The analysis of the twelve best measured cases indicates that such agreement occurs at the level of a few parts in $10^{4}$\cite{Towner:2005qc}. Since the unitarity deficit had been at the part per thousand level, it is hard to see how nuclear structure effects could be the reason in light of the almost order of magnitude more precise agreement with CVC among the twelve superallowed cases of interest. 

Recently, new measurements of $K_{e3}$ decays have lead to a shift in the world average for the branching ratios (for a recent summary, see Ref.~\cite{Blucher:2005dc}). The extraction of $V_{us}$ from these branching ratios, however, depends on knowledge of a $K$-to-$\pi$ form factor, $f_{+}(q^2=0)$, and there currently exists some controversy over the value of this quantity. It can be analyzed in chiral perturbation theory (ChPT) and all contributions through order $p^4$ are known. These contributions include both one-loop corrections and contributions from tree-level ${\cal O}(p^4)$ operators whose coefficients are known from other experiments. To perform a test of the CKM matrix at the 0.1\% level, however, the ${\cal O}(p^6)$ contributions are also required. The loop contributions (that include both one- and two-loop corrections) have been computed\cite{Post:2001si,Bijnens:2003uy}, but the tree-level ${\cal O}(p^6)$ operator contributions are not known in a model-independent way. Values for these quantities have been obtained using large-$N_C$ QCD methods\cite{Cirigliano:2005xn} and lattice QCD computations\cite{Becirevic:2004ya,Okamoto:2004df,Tsutsui:2005cj,Dawson:2005zv}. The results of the two approaches point in contradictory directions: the large-$N_C$ value for $f_{+}(0)$ would yield a value for $V_{us}$ that -- in combination with the nuclear result for $V_{ud}$ -- disagrees with unitarity at the historically irritating $\sim 2\sigma$ level; the quenched lattice results, in contrast, point to unitarity agreement. It will be quite interesting to see how this situation settles down in the next few years as hadron structure calculations improve. In the meantime, new tests of the nuclear structure corrections that are applied to the experimental $ft$ values are being carried out in regions of the periodic table where they are expected to be larger. Future studies at a rare isotope facility would presumably be of interest in this respect. 

The implications of unitarity tests for new physics at the weak scale can best be understood by considering the relationship between the vector Fermi constant, $G_V^\beta$,  that governs nuclear decays with the Fermi constant, $G_\mu$,  obtained from the muon lifetime. One has  
\be
\label{eq:gvbeta}
G_V^\beta=G_\mu\, V_{ud}\, \left(1+{ \Delta\hat r}_\beta-{ \Delta \hat r}_\mu\right) \ \ \ ,
\ee
where ${ \Delta \hat r}_{\beta,\mu}$ are ${\cal O}(\alpha/4\pi)$  electroweak radiative corrections to the two decay amplitudes. The quantity ${ \Delta \hat r}_\beta$ contain hadronic structure uncertainties associated with the $W\gamma$ box graphs involving the lepton and nucleon pairs. Recently, Marciano and Sirlin have reduced this uncertainty by relating the short distance part of the one-loop integral to the Bjorken sum rule and by applying large $N_C$ correlators to the pertrubative-nonperturbative transition region\cite{Marciano:2005ec}. The resulting value for $V_{ud}$ obtained by these authors and by an up-dated global analysis of superallowed decays\cite{Hardy:2005qv} is
$ V_{ud}=0.97377(11)(15)(19)$, where the first error is the combined experimental $ft$ error and theoretical nuclear structure uncertianty; the second is associated with nuclear coulomb distortion effects; and the last error is the theoretical hadronic structure uncertainty.

New physics can contribute to ${\Delta \hat r}_\beta-{ \Delta \hat r}_\mu$ either at tree-level or through loop effects. In SUSY, for example, tree-level effects arise if one allows for terms that violate lepton number\cite{Barger:1989rk,Ramsey-Musolf:2000qn}. Such interactions can be completely supersymmetric and not forbidden by any other symmetry principle\footnote{The analogous terms that violate baryon number must be vanishingly small, however, in order to avoid proton decay that is too rapid.}. A CKM unitarity deviation of any sign could be compatible with the presence of such \lq\lq R parity" violating interactions, so long as one permits effects in both $\mu$ and $\beta$-decays simultaneously\cite{Ramsey-Musolf:2000qn}. If one imposes R parity conservation, supersymmetric corrections arise at loop level. My former student A. Kurylov and I analyzed these corrections and found that a unitarity deviation could have interesting implications for the spectrum of supersymmetric particles and models people construct that try to explain why these particles are heavier than the SM particles\cite{Kurylov:2001zx}. From this perspective, then, it is important to know whether or not the CKM matrix is unitarity. 

\vskip 0.1in

\noindent{\em $\beta$-decay correlations.} Another interesting tool for probing new weak scale physics with $\beta$-decay is to study the spectral shape, spatial distribution, and polarization of the outgoing $\beta$ particles. These characteristics of the spectrum are described by various correlations, as noted many years ago by Jackson, Treiman, and Wyld\cite{jtw} ( for recent discussions, see also \cite{Herczeg:2001vk,deutsch,Severijns:2006dr}) . For our purposes, it is useful to write the partial rate as 
\begin{eqnarray}
\label{eq:betacor}
   d\Gamma& \propto & {\cal N}(E_e)\Biggl\{ 1+a {{\vec p}_e\cdot{\vec p}_\nu\over E_e E_\nu}
   + b{\Gamma m_e\over E_e} + \langle {\vec J}\rangle\cdot \left[A{{\vec p}_e\over E_e} 
   + B{{\vec p}_\nu \over E_\nu} + D{{\vec p}_e\times {\vec p}_\nu \over E_e E_\nu}\right] \\
 \nonumber
&&+ {\vec\sigma}\cdot\left[N \langle{\vec J}\rangle + G\frac{{\vec p}_e}{E_e}+Q^\prime {\hat p}_e {\hat p}_e\cdot \langle{\vec J}\rangle+R \langle {\vec J}\rangle\times\frac{{\vec p}_e}{E_e}\right]
 \Biggr\}
   d\Omega_e d\Omega_\nu d E_e,
\end{eqnarray}
where ${\cal N}(E_e)=p_e E_e(E_0-E_e)^2$; $E_e$ ($E_\nu$), ${\vec p}_e$ 
(${\vec p}_\nu$), and ${\vec\sigma}$ are the $\beta$ (neutrino) energy, momentum, and polarization, respectively; ${\vec J}$ is the  polarization of the decaying nucleus; and $\Gamma=\sqrt{1-(Z\alpha)^2}$. 

The various coefficients of the various correlations in this expression are sensitive to low-energy CC interactions that differ from the $(V-A)\otimes(V-A)$ structure of the SM. It is helpful to characterize these interactions using an effective four fermion Lagragian\cite{Profumo:2006yu}
\begin{equation}
\label{eq:leffbeta}
{\cal L}^{\beta-\rm decay} = - \frac{4 G_\mu}{\sqrt{2}}\ \sum_{\gamma,\, \epsilon,\, \delta} \ a^\gamma_{\epsilon\delta}\, 
\ {\bar e}_\epsilon \Gamma^\gamma \nu_e\, {\bar u} \Gamma_\gamma d_\delta
\end{equation}
where the sum is over all Dirac matrices $\Gamma^\gamma= 1$ (S), $\gamma^\alpha$ (V), and $\sigma^{\alpha\beta}/\sqrt{2}$ (T) \footnote{The normalization of the tensor terms corresponds to the convention adopted in Ref.~\cite{Scheck}}. and fermion chiralities ($\epsilon$,$\delta$). Within the SM, only $a^V_{LL}$ is non-vanishing, and in this case one has 
\be
a^V_{LL}=V_{ud}\, \left(1+{\Delta\hat r}_\beta-{ \Delta\hat r}_\mu\right)\ \ \ ,
\ee
in accordance with Eq.~(\ref{eq:gvbeta}) above. 

Contributions to some of the other $a^\gamma_{\epsilon\delta}$ that arise at tree-level in models with RH gauge bosons or leptoquarks have been reviewed extensively by Herczeg\cite{Herczeg:2001vk}, and I refer the reader to his work for further details. Here, I wish to summarize a recent analysis of non-$(V-A)\otimes(V-A)$ $\beta$-decay operators induced at one-loop order in SUSY\cite{Profumo:2006yu}. These effects arise when the scalar superpartners of left- and right-handed fermions mix through interactions that break supersymmetry known as triscalar interactions. If ${\tilde f}_{L,R}$  denote these scalar fermions then their mass matrix 
\be
{\bf M_{\tilde f}^2} =\left(
\begin{array}{cc}
{\bf M_{LL}^2} & {\bf M_{LR}^2}\\
{\bf M_{LR}^2} & {\bf M_{RR}^2}
\end{array}\right)
\ee
will lead to L-R mixing when the off-diagonal entry is non-zero\footnote{In principle, the matrix can also mix superpartners of different generations, so ${\bf M_{AB}^2}$ are $3\times 3$ matrices in flavor space.}. In the MSSM this term is 
\begin{equation}
\label{eq:mlr}
{\bf M_{LR}^2}={\bf M_{RL}^2} = \biggl\{
\begin{array}{cc}
v\left[{\bf a_f} \sin\beta -\mu {\bf Y_f} \cos\beta\right]\ , & {\tilde u}-{\rm type\ sfermion}\\
v\left[{\bf a_f} \cos\beta -\mu {\bf Y_f} \sin\beta\right]\ , & {\tilde d}-{\rm type\ sfermion}
\end{array}\ \ \ .
\end{equation}
The MSSM contains two Higgs doublets and  $\tan\beta=v_u/v_d$ gives the ratio of the vacuum expectation value of their  neutral components, with $v=\sqrt{v_u^2+v_d^2}$. The quantity $\mu$ is a dimensionful parameter that characterizes a supersymmetric coupling between the two Higgs doublets. The matrices  ${\bf Y_f}$ and ${\bf a_f}$ are the $3\times 3$ Yukawa and soft triscalar couplings. The presence of the latter breaks supersymmetry, and in many models, it is often assumed to be proportional to ${\bf Y_f}$. Under this \lq\lq alignment"  assumption, the off-diagonal term ${\bf M_{LR}^2}$ is suppressed relative to the diagonal terms in the mass matrix for the first and second generation sfermions. 

It would be interesting to test alignment assumption for the first two generations, yet doing so with collider studies could be difficult. The study of $\beta$-decay correlations, in contrast, may provide a targeted means for carrying out such a test. If the L-R mixing terms are not Yukawa-suppressed and if the mixing is close to maximal, then one-loop effects involving L-R mixed sfermions would induce scalar and tensor operators in ${\cal L}^{\beta-\rm decay}$. The resulting effects in the partial rate would appear in the Fierz interference coefficient, $b$; the energy-dependent components of the neutrino asymmetry parameter, $B$, and spin-polarization correlation coefficient $Q^\prime$; and the energy-independent term in the spin-polarization correlation $N$. Superallowed nuclear decays, for example, are sensitive to the scalar contribution to $b$. In the limit that we neglect the chargino ($\tilde\chi^\pm$) and neutralino ($\tilde\chi^0$) mixing matrices, we have 
\be
\label{eq:bf}
b_F \approx \frac{2\alpha}{3\pi}\,  \left(\frac{g_S}{g_V}\right) \, {\rm Re}\, Z_L^{1m} Z_L^{4 m\ast} \left[ \left(Z^{1i\ast}_D Z^{4i}_D\right)\, M_Z^2\, {\cal F}_1- \left(Z^{1i\ast}_U Z^{4i}_U\right)\, M_Z^2\, M_{\tilde\chi^+} M_{\tilde\chi^0}\,  {\cal F}_2\right]\ \ \ .
\ee
Here, $g_V$ and $g_S$ are vector and scalar form factors of the nucleon; ${\cal F}_{1,2}$ are loop functions of the superpartner masses; and the $Z_F^{ij}$ are elements of the matrices that rotate the sfermion weak eigenstates into corresponding mass eigenstates for superpartners of the charged leptons ($L$), down ($D$), and up ($U$) quarks. The combinations $Z_F^{1m} Z_F^{4m\ast}$ appearing in Eq.~(\ref{eq:bf}) are non-zero only in the presence of L-R mixing among sfermionic superpartners of first generation fermions $F$. 

How precisely would studies of the $\beta$ spectrum need to be in order to probe these effects at an interesting level? For superpartner masses of order the electroweak scale, the products of $M_Z^2$ and the loop functions can be as large as ${\cal O}(10^{-1})$; the prefactor $2\alpha/3\pi$ is $\lsim{\cal O}(10^{-2})$; so for large L-R mixing for which the products $Z_F^{1m} Z_F^{4m\ast}$ are ${\cal O}(1)$ the net effect on $b_F$ can approach $10^{-3}$. The present limits on $b_F$ are 0.0026(26). With substantial improvements in sensitivity, measurements of the $\beta$ spectrum that probe this term could search for large L-R mixing and test the alignment hypothesis. Determinations of the correlation coefficients with similar levels of sensitivity would serve the same purpose. Future measurements of the energy-dependence of the neutrino asymmetry parameter $B$ at the few $\times 10^{-4}$ level appear feasible using cold or ultracold neutrons at the Spallation Neutron Source. In either case, null results at this level would point to L-R mixing that is substantially non-maximal and lend experimental credence to the alignment hypothesis. 

\section{Weak Interactions of Quarks}

The weak interactions of quarks at low energies remain enigmatic, despite decades of theoretical and experimental scrutiny. While we have no strong reasons to doubt the the SM prediction for the structure of the elementary weak quark-quark interaction, we have only a limited grasp of the ways in which this interaction becomes manifest in strongly-interacting systems where nonperturbative QCD affects the weak interaction dynamics. In the purely strong interaction sector, the use of symmetries such as chiral, heavy quark, large $N_C$, and heavy quark symmetries -- and the effective theories built on them -- have given us powerful tools for explaining strong interaction dynamics without having to compute them from first principles in QCD. Ideally, similar methods would help us explain the hadronic weak interaction (HWI), but our experience is thus far not encouraging. In the $\Delta S=1$ sector, for example, one cannot simultaneously account for the parity conserving and parity violating (PV) nonleptonic decays of hyperons using chiral symmetry. Similarly, the radiative decays of hyperons produce significantly larger PV asymmetries than one would expect based on the breaking of vector SU(3) symmetry by the strange quark mass. Even the well-known $\Delta I=1/2$ rule defies explanation based on symmetry considerations. 

It may be that our inability to apply successfully the approximate symmetries of QCD to the $\Delta S=1$ HWI results from the presence of the strange quark with its problematic mass of order the QCD scale. Or, it may be that we are missing a more fundamental aspect of the dynamics involving the interplay of weak and strong interactions of quarks that apply to both light quarks and strange quarks. In order to find out, it is useful to study the $\Delta S=0$ HWI which is largely devoid of strange quark effects. The difficulty, however, is that we can experimentally access only the PV part of this interaction, since the strong and electromagnetic interactions are considerably larger than the parity conserving $\Delta S=0$ HWI at low energies. 

Historically, the PV $\Delta S=0$ HWI has been probed using a combination of polarized proton scattering experiments and observation of PV processes in nuclei. The latter have been quite attractive  since accidents of nuclear structure can amplify the PV effects making them easier to access experimentally. Unfortunately, the theoretical interpretation of nuclear observables is complicated by nuclear structure, and to date, we have not obtained a consistent description of the $\Delta S=0$ HWI using nuclear probes. As people at this meeting are well aware, the standard framework for describing this interaction in nuclei has been a meson-exchange model, popularized by Desplanques, Donoghue, and Holstein. The model contains seven PV meson-nucleon couplings, $h_M^{i}$, and different nuclear experiments are sensitive to different combinations of these coupling. One such coupling, $h_\pi^1$, has received considerable attention, since it is the only PV pion-nucleon coupling and, therefore, the only one that parameterizes the strength of the long-range PV force between nucleons. Measurements of the PV $\gamma$-decays of $^{18}F$ imply that this coupling is consistent with zero and that there is no appreciable, long-range component to the $\Delta S=0$ HWI. On the other hand, the results for the anapole moment of $^{133}$Cs obtained in atomic PV by the Boulder group imply the presence of a large $h_\pi^1$. Unless the cesium result is an aberration, there is clearly something about the HWI in nuclei that we do not understand (for a recent discussion of these and other issues, see Ref.~\cite{Ramsey-Musolf:2006dz}).

Fortunately, there exists a way forward. Experimentally, new techniques involving few-body systems make measurements of ${\cal O}(10^{-7})$ PV effects in this arena feasible where there were not two decades ago. Theoretically, the impressive developments using Green's function and variational Monte Carlo methods make it possible to perform {\em ab initio} computations in few-body systems starting from a given nucleon-nucleon potential. In addition, the framework for describing the $\Delta S=0$ HWI has been reformulated using effective field theory (EFT), thereby allowing one to circumvent the untestable model-dependence of the meson-exchange framework\cite{Zhu:2004vw}. 
What has emerged from this confluence of developments is the potential of a new program of few-body hadronic PV studies that could determine the  PV \lq\lq low-energy constants" of the corresponding EFT to lowest order in a way that is free from nuclear structure and hadronic model uncertainties. Completion of this program will require (i) new measurements with neutrons and few-body nuclei at facilities such as the SNS, and (ii) new theoretical computations of the corresponding observables using the EFT framework. I am optimistic that completion of this program will lead to a deeper understanding of the $\Delta S=0$ HWI  and shed new light on the long standing puzzles in the $\Delta S=1$ sector.

Why is such a program of interest to a future rare isotope facility? The answer is that it could help us better understand the nucleus as a \lq\lq laboratory" for searching for new physics. At the most basic level, the low-energy HWI involves nuclear and hadronic matrix elements of four quark operators. The program outlined above will help us better understand the dynamics of such matrix elements in the few-nucleon system. Moreover, if the program is successful, we will have in hand a well-determined, PV NN potential to ${\cal O}(p)$. This potential could then be used to compute nuclear PV observables, such as the $^{18}$F $\gamma$-decay polarization or anapole moments of complex nuclei like cesium or francium that could be studied with a rare isotope facility. A comparison of such computations and experimental results will teach us whether or not an EFT weak interaction potential can adequately describe the dynamics of four-quark operators in the nuclear environment -- rather than trying to use nuclear observables to determine the potential in a model-dependent way.

In the best of all possible worlds, the EFT approach to treating four-quark operators in nuclei will be successful, thereby allowing us to make progress in the interpretation of another important class of studies where four-quark operators can contribute: neutrinoless double $\beta$-decay ($0\nu\beta\beta$). The most important aspect of $0\nu\beta\beta$ is that it may tell us that neutrinos are Majorana fermions. However, people in the business would also like to use it to determine the absolute scale of neutrino mass. This second use of $0\nu\beta\beta$ makes sense so long as there exist no competing contributions to the decay rate from heavy particles with lepton number violating interactions. Unfortunately, reasonable candidates for such particles exist -- such as the scalar fermions of SUSY when R parity is not conserved. If the mass scale of such particles is not too different from the electroweak scale, then their contributions to the $0\nu\beta\beta$ rate can be comparable to those from the exchange of a light Majorana neutrino\cite{Cirigliano:2004tc}. In this case, one would need to compute the effects of heavy particle exchange  in order to separate them out from the possible light Majorana neutrino exchange, and doing so requires calculating nuclear matrix elements of the four-quark operators generated by heavy particle exchange . If our EFT methods for the HWI in nuclei are successful for that problem, then the same techniques could be used with some confidence in the case of heavy particle contributions to $0\nu\beta\beta$, thereby sharpening the interpretation of these important nuclear weak decay experiments. To this end, an EFT formulation for heavy particle contributions to $0\nu\beta\beta$ has recently been developed\cite{Prezeau:2003xn}.

\section{Conclusions}

The primary aim of a new  rare isotope facility will, of course, be to to study novel aspects of nuclear structure that are not accessible with present facilities. Carrying out such studies will be an important part of the overall mission of nuclear physics. At the same time, such a facility will provide new opportunities to use nuclei as laboratories to probe the dynamics of quarks at times that preceded the confinement era, to look for evidence of new forces that affected quarks during early times, and to sort out the properties of neutrinos. I hope that in this talk I have conveyed a sense of the opportunities at hand with such a facility -- opportunities that I hope our community will one day realize.


\vspace*{1cm}
\noindent{{\bf Acknowledgments} } \\
\noindent 
I wish to thank the organizers of the workshop for financial support and hospitality.
This work was supported in part under U.S. DOE contract
DE-FG02-05ER41361 and NSF grant PHY-0555674.

\end{document}